\documentclass[%
 aip,
 amsmath,amssymb,
 reprint,%
]{revtex4-2}

\usepackage{graphicx}% Include figure files
\usepackage{dcolumn}% Align table columns on decimal point
\usepackage{bm}% bold math
\usepackage{calrsfs}
\usepackage[utf8]{inputenc}
\usepackage[T1]{fontenc}
\usepackage{mathptmx}
\usepackage{etoolbox}
\DeclareMathAlphabet{\pazocal}{OMS}{zplm}{m}{n}
\makeatletter
\def\@email#1#2{%
 \endgroup
 \patchcmd{\titleblock@produce}
  {\frontmatter@RRAPformat}
  {\frontmatter@RRAPformat{\produce@RRAP{*#1\href{mailto:#2}{#2}}}\frontmatter@RRAPformat}
  {}{}
}%
\makeatother
\begin{document}

% \preprint{AIP/123-QED}

\title[]{Chimera states emerging from dynamical trapping in chaotic saddles}

\author{Everton S. Medeiros}
\email{everton.medeiros@uni-oldenburg.de}
\affiliation{Institute for Chemistry and Biology of the Marine Environment, Carl von Ossietzky University Oldenburg, 26111 Oldenburg, Germany}

\author{Oleh Omel'chenko}
\affiliation{Institute of Physics and Astronomy, University of Potsdam, Karl-Liebknecht-Str. 24/25, 14476 Potsdam, Germany}

\author{Ulrike Feudel}
\affiliation{Institute for Chemistry and Biology of the Marine Environment, Carl von Ossietzky University Oldenburg, 26111 Oldenburg, Germany}

\date{\today}

\begin{abstract}
Nonlinear systems possessing nonattracting chaotic sets, such as chaotic saddles, embedded in their state space may oscillate chaotically for a transient time before eventually transitioning into some stable attractor. We show that these systems, when networked with nonlocal coupling in a ring, are capable of forming chimera states, in which one subset of the units oscillates periodically in a synchronized state forming the coherent domain, while the complementary subset oscillates chaotically in the neighborhood of the chaotic saddle constituting the incoherent domain. We find two distinct transient chimera states distinguished by their abrupt or gradual termination. We analyze the lifetime of both chimera states, unraveling their dependence on coupling range and size. We find an optimal value for the coupling range yielding the longest lifetime for the chimera states. Moreover, we implement transversal stability analysis to demonstrate that the synchronized state is asymptotically stable for network configurations studied here.
\end{abstract}

\maketitle

\begin{quotation}
We consider ring networks of identical dynamical units whose behavior is determined by a globally stable periodic orbit and a nonattracting chaotic set (e.g., a chaotic saddle) in their state space. In the presence of a finite range of nearest-neighbor coupling, such networks develop an effective bistability characterized by a fully synchronized periodic state that stably coexists with a chaotic desynchronized one. These networked systems are extremely sensitive to perturbations of arbitrary size which may lead to the formation of a transient chimera state, i.e., a state where only a portion of the network nodes synchronizes, while others evolve asynchronously near the nonattracting chaotic set. We perform a thorough analysis of such chimera states and report a number of their properties that differ from those of usual chimera states found in networks of coupled oscillators and coupled map lattices.
\end{quotation}

\section{Introduction}
Many natural and man-made systems are modeled as networks of interacting dynamical units. In the presence of non-local coupling (e.g., finite-range and/or distance-dependent coupling) such networks can exhibit complex patchy dynamics even when all individual units are identical. The most prominent examples of such spatiotemporal patterns are {\it chimera states} in networks of coupled oscillators~\cite{kurbat02,abrstr04} and {\it bump states} in neural networks~\cite{lc01,lai15,sa20}. Although the occurrence of both these states is determined by similar dynamical mechanisms~\cite{fow21}, their appearance is slightly different due to the different nature of individual dynamical units.

Since their discovery by Kuramoto and Battogtokh in 2002, chimera states have been the subject of intensive research~\cite{pa2015,Sch2016,KemHSKK2016,o2018,pjaswbp2021}, which unveiled a number of surprising features. In particular, it was shown that the dynamics of chimera states is associated with weak extensive chaos~\cite{woym11}. Moreover, it was found that the chimera states often behave not as attractors but as long-lived chaotic transients~\cite{wo11}. In addition, for the chimera states on the ring networks it was observed that their positions are not fixed, but exhibit erratic Brownian motion~\cite{owm10}. Although most of the above facts were first established for chimera states in networks of phase oscillators, their relevance was later also demonstrated in more realistic systems consisting of limit-cycle oscillators~\cite{Sch2016}, in coupled map lattices~\cite{omhs11}, and even in networks of coupled chaotic units~\cite{gsd13}. Moreover, following these theoretical and numerical works, the existence of chimera states was confirmed in laboratory experiments with chemical~\cite{tns2012,nts2013,TotRTSE2018}, electrochemical~\cite{WicK2013,SchSKG-M2014}, mechanical~\cite{mtfh2013,KapKWCM2014} and optoelectronic~\cite{hmrhos2012} oscillators.

A chimera state is usually characterized as a pattern formation phenomenon in which a symmetrically coupled network of identical units spontaneously splits into two groups with synchronized and desynchronized dynamics. In the synchronized group, all individual units stay near their common attractor, whereas the desynchronized group units approach the attractor only occasionally, taking large detours around it for the rest of the time. Another mechanism supporting the formation of chimera states is found in the case of bistable dynamical units with two coexisting qualitatively different attractors (e.g., a fixed point and a chaotic attractor)~\cite{dmk2014,dmk2016,mhbrd15,ccfg-nr16}. Synchronized and desynchronized groups are then identified by the fact near which attractor the unit moves. In all these cases, as we can see, the key role in the emergence of chimera states is played by the attractor of the individual dynamical unit, while the global organization of its state space is less important. However, this is not a general rule. In this paper, we describe chimera states that arise through a qualitatively different mechanism in which both of the above factors are crucial. More specifically, we consider nonlocally coupled networks of identical oscillators, each one possessing chaotic saddles coexisting with periodic attractors in their individual state space. Recall that a chaotic saddle is an invariant chaotic set of a nonlinear dynamical system with stable and unstable directions, along which the trajectories of the system are attracted to it or repelled from it. Chaotic saddles occur in various applications and play an important role there
\cite{Skufca2006,Eckhardt2007,Rempel2007,Joglekar2015,Ansmann2016,lilienkamp2017,Lucarini2019}.
In particular, the combination of a chaotic saddle and a periodic attractor in the state space leads to the fact that the corresponding dynamical system behaves as a chaotic transient~\cite{Lai2011} or supertransient~\cite{Crutchfield1988, Lai1995,Lai2011}. Recently, it has been found that networked systems possessing chaotic saddles in their state space may give rise to extraordinarily long chaotic transients or even chaotic attractors \cite{Medeiros2018,Medeiros2019,Medeiros2021}, when network trajectories oscillate chaotically for times indefinitely long in a completely desynchronized configuration. It turns out that this completely desynchronized state coexists with the completely synchronized one in the network's state space. Moreover, if we perturb just one oscillator in the synchronized state, a certain portion of its neighbors usually follows this oscillator, leaving the synchronized state and approaching the nonattracting chaotic set.

In this paper, we show that the perturbations to the synchronized state in the aforementioned networked systems lead, in fact, to a novel type of chimera state. More specifically, we thoroughly investigate such chimera states, focusing mainly on their chaotic transient nature. The paper is organized as follows. First, we describe our main model --- a nonlocally coupled network of Duffing oscillators and perform a stability analysis of its synchronization manifold. Next, we show how transient chimera states emerge in the system due to a small perturbation of only one oscillator. We find that, in contrast to other known chimera states, the chimera states in our system decay in two different directions to either a completely synchronized state or a homogeneous desynchronized state. We call them the $S$-type and $D$-type chimeras, respectively. The $S$-type chimeras terminate abruptly, similar to the so-called type-II chaotic transients~\cite{Lai2011}. In contrast, the decay of $D$-type chimeras resembles a gradual (although nonmonotonic) process when the desynchronized ``phase'' invades into the synchronized one. We perform statistical analysis of the lifetimes of both types of chimera states for different coupling ranges and sizes. Moreover, we demonstrate the generality of our findings by reproducing them in the network of coupled logistic maps. Finally, we summarize the obtained results in the conclusions section.

\section{Ring network of Duffing oscillators}

The high-dimensional equations describing the dynamics of $N$ coupled oscillators are given by:
\begin{eqnarray}
 \dot{{\bf r}}_i &=& {\bf F}({\bf r}_i)+\sigma \sum_{j=1}^{N}G_{ij}{\bf H}({\bf r}_j).
 \label{Eq:gen_model}
\end{eqnarray}
where the vectors $\mathbf{r}_i=(x_i,y_i)$, with $i=1,\dots,N$, specify the state space of each oscillator in the network. The function $\bf{F}(\bf{r})$ prescribes the nonlinear dynamics of each oscillator, while ${\bf H}({\bf r})$ stands for the coupling function. The constant $\sigma$ controls the coupling intensity among the oscillators. Finally, the connectivity matrix ${\bf G}$ specifies the coupling architecture.

Now, we specify our network's dynamical features and coupling structure. The results reported here occur for a wide range of dynamics at the network nodes, only requiring two main ingredients of the oscillators: a stable periodic attractor and a chaotic saddle. We initially chose the Duffing oscillator to meet this generic requirement. The equation of motion of this oscillator is written as follows:
\begin{equation}
\mathbf{F}(\mathbf{r})=
 \begin{pmatrix}
 y \\
-\gamma y + x - x^3 + A \cos(\omega t)
\end{pmatrix},
\label{Eq:dynamical}
\end{equation}
where $\mathbf{r}=(x,y)$ specifies the coordinates of the non-autonomous Duffing oscillator. The parameter $\gamma$ controls the energy dissipation in the oscillators. The parameters $A$ and $\omega$ are the external forcing's amplitude and frequency, respectively. The coupling among the Duffing oscillators is specified by a linear (diffusive) coupling function $\mathbf{H}$ of the form:
\begin{equation}
\mathbf{H}(\mathbf{r})=
 \begin{pmatrix}
x \\
y
\end{pmatrix}.
\label{Eq:engage}
\end{equation}
With this coupling function, the Duffing oscillators in our network interact with each other through both components $(x,y)$ of their equations. In addition, the coupling structure adopted in Eq.~(\ref{Eq:gen_model}) is captured by a $N \times N$ connectivity matrix $\mathbf{G}$ as:
\begin{equation}
{\footnotesize
%\nonumber
\mathbf{G}=
 \begin{pmatrix}
-2R & 1 & \cdots & 1 & 0 & \cdots & 0 & 1 & \cdots & 1 \\
1 & -2R & 1 & \cdots & 1 & 0 & \cdots & 0 & 1 & \cdots \\
\vdots & \vdots & \vdots & \ddots & \vdots & \vdots & \ddots & \vdots & \vdots & \vdots \\
\cdots & 1 & 0 & \cdots & 0 & 1 & \cdots & 1 & -2R & 1\\
1 & \cdots & 1 & 0 & \cdots & 0 & 1 & \cdots & 1 & -2R
\end{pmatrix}.
}
\label{Eq:connectivity}
\end{equation}
This cyclic connectivity matrix describes a nonlocal coupling structure with periodic boundary conditions forming a ring topology. The constant $R$ specifies the coupling range of each oscillator, i.e., the number of neighbor oscillators $j$ coupled to oscillator $i$ on each side. In addition, the matrix ${\bf G}$ satisfies the condition $\sum_{j=0}^{N}G_{ij}=0$ for any $i$ and it can be diagonalized with a set of eigenvalues $0=\lambda_1 > \lambda_2 \geqslant \lambda_3 \geqslant \dots \geqslant \lambda_N$.

Finally, to obtain an explicit form for the equations describing a network of Duffing oscillators, we insert Eqs.~(\ref{Eq:dynamical})--(\ref{Eq:connectivity}) into Eq.~(\ref{Eq:gen_model}).
In addition, we normalize the coupling intensity $\sigma$ by the connectivity of the oscillators $2R$. With this, we obtain the network equations in an explicit form:
\begin{eqnarray}
 \dot{x}_{i} &=& y_{i} + \frac{\sigma}{2R} \sum\limits_{j=i-R}^{j=i+R} ( x_{j} - x_{i} ),
 \label{Eq:x_explicit}\\[1mm]
 \nonumber
 \dot{y}_i &=& -\gamma y_i + x_i - x_i^3 + A \cos(\omega t) + \frac{\sigma}{2R} \sum\limits_{j=i-R}^{j=i+R} ( y_j - y_i ),
 \label{Eq:y_explicit}
\end{eqnarray}
where the variables $(x_i,y_i)$ specify the state space of the network composed of Duffing oscillators.

\begin{figure}[!htp]
\centering
\includegraphics[width=8.5cm,height=4cm]{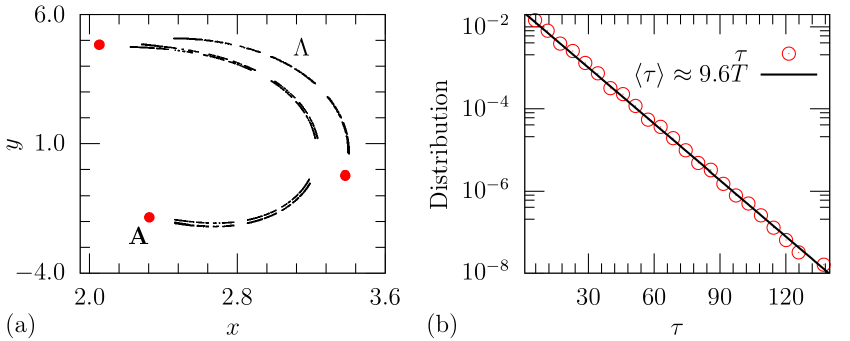}
\caption{(a) State space of the isolated Duffing oscillator ($\sigma=0$). The period-$3$ attractor $\mathbf{A}$ is represented by the red circles, while the black dots display an approximation of the chaotic saddle $\Lambda$. (b) The red circles indicate the normalized distribution of escape times from the chaotic saddle. The black curve is an exponential fit providing the mean escape time $\langle \tau \rangle = 9.6T$.}
\label{Fig:figure_1}
\end{figure}

Next, before we explore the dynamics exhibited by the network in Eq.~(\ref{Eq:x_explicit}), we examine the state space of the uncoupled Duffing oscillator ($\sigma=0$). Recalling that the basic requirements for the oscillator's dynamics consist in the coexistence of a chaotic saddle and stable periodic attractor, we fix the parameters of the Duffing oscillator at $\gamma=0.24$, $A=13.6330$, and $\omega=0.5$. For these parameters, a chaotic saddle and a globally stable limit cycle occur in the state space of the oscillator. To visualize this configuration, we define a stroboscopic Poincar\'e section in which the variables $(x,y)$ are sampled at every period $T=2\pi/\omega$ of the oscillator forcing. In this section, the stable limit cycle is a period-$3$ attractor, $\mathbf{A}$, [red circles in Fig.~(\ref{Fig:figure_1})(a)], while the chaotic saddle $\Lambda$ forms a fractal set [black dots in Fig.~(\ref{Fig:figure_1})(a)]. Once the attractor $\mathbf{A}$ is globally stable, all trajectories ultimately approach it regardless of their initial condition. However, a set of trajectories will detour through the chaotic saddle, causing them to oscillate chaotically for a finite time (chaotic transient) before escaping to the attractor $\mathbf{A}$. The escape times from the chaotic saddle are known to be exponentially distributed \cite{Lai2011}. We measure the escape times $\tau$ of the chaotic saddle $\Lambda$ for different initial conditions and denote it by red circles in Fig.~(\ref{Fig:figure_1})(b). By fitting a normalized exponential distribution $\rho(\tau) =\rho_0 e^{-\tau/\langle \tau \rangle}$ to the measured escape times [see black line in Fig.~(\ref{Fig:figure_1})(b)], we obtain the mean escape time of $\Lambda$ as $\langle \tau \rangle=9.6T$.

\section{Linear and global stability of the synchronization manifold}

Following the understanding of the local dynamics of each Duffing oscillator in the previous section, we move to the solutions of the network in Eq.~(\ref{Eq:x_explicit}). The first network solution of interest is the completely synchronized state specified by the dynamics at a synchronization manifold $\mathbf{s}=\mathbf{r}_1=\mathbf{r}_2= \cdots = \mathbf{r}_N$. Since the completely synchronized state $\mathbf{s}$ is a solution emerging from the interactions among the oscillator, it naturally depends strongly on the network characteristics, i.e., the coupling intensity $\sigma$, the size $N$, and the coupling range $R$. With this in mind, we now analyze the linear and global stability of the completely synchronized state in the network (Eq.~\ref{Eq:x_explicit}) with each oscillator containing the periodic attractor $\mathbf{A}$ and the chaotic saddle $\Lambda$.

\begin{figure}[!htp]
\centering
\includegraphics[width=7cm,height=5cm]{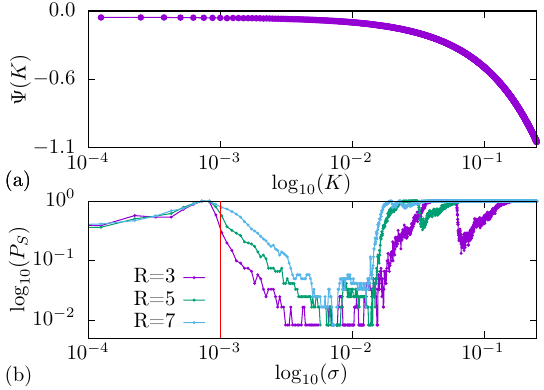}
\caption{(a) Master stability function $\Psi(K)$ obtained as a function of the $K=\sigma |\lambda_i|$. (b) Probability of synchronization $P_S$ as a function of the coupling intensity $\sigma$. The network size is fixed at $N=25$.}
\label{Fig:figure_2}
\end{figure}

\textit{Linear stability:} In the study of synchronization, a natural concern is the asymptotic stability of the completely synchronized state $\mathbf{s}$. This information is particularly important when the oscillators composing the network are chaotic. In our case, the oscillators oscillate periodically in the attractor $\mathbf{A}$. However, the occurrence of transient chaos is also a reason for questioning the asymptotic stability of the synchronized behavior. To address this issue, we implement the formalism of the master stability function \cite{Pecora2000} for our network. For that, we consider the network equation in the form of Eq.~(\ref{Eq:gen_model}) to apply infinitesimal perturbations $\delta {\bf r}_i$ to all state space directions. The variational equation describing the time evolution of the perturbations is given by:
\begin{eqnarray}
 \delta\dot{{\bf r}}_i &=& {\bf DF}({\bf s})\cdot\delta{\bf r}_i +\sigma\sum_{j=1}^{N}G_{ij}{\bf DH}({\bf s})\cdot\delta{\bf r}_j,
 \label{Eq:variational}
\end{eqnarray}
where $\mathbf{DF}({\bf s})$ and $\mathbf{DH}({\bf s})$ are Jacobian matrices of the respective functions in Eq.~(\ref{Eq:dynamical}) and Eq.~(\ref{Eq:engage}) evaluated at ${\bf s}$. Since the matrix ${\bf G}$ can be diagonalized with a set of real eigenvalues $\{\lambda_i,i=1,\dots,N\}$, the second term of Eq.~(\ref{Eq:variational}) is block diagonalized, resulting in a decoupled variational equation:
\begin{eqnarray}
 \delta\dot{{\bf \zeta}}_i &=& \left[{\bf DF}({\bf s})-\sigma|\lambda_i|{\bf DH({\bf s})} \right]\cdot\delta{\bf \zeta}_i.
 \label{Eq:diagonalized}
\end{eqnarray}
Substituting $\sigma |\lambda_i|=K$ in Eq. (\ref{Eq:diagonalized}), a generic variational equation is written as:
\begin{eqnarray}
\delta\dot{{\bf \zeta}} &=& \left[{\bf DF}({\bf s})-K{\bf DH({\bf s})}\right]\cdot\delta{\bf \zeta}.
\label{Eq:generic}
\end{eqnarray}
The largest nonzero Lyapunov exponent of Eq. (\ref{Eq:generic}) yields the MSF $\Psi(K)$. Hence, in Fig. \ref{Fig:figure_2}(a), we obtain $\Psi(K)$ as a function of $K$. We observe that $\Psi(K)<0$ for all $K>0$. This result indicates that the completely synchronized state is asymptotically stable for any coupling intensity and architecture when the considered Duffing oscillator is coupled through both of its components.

\textit{Global stability:} This type of stability refers to a solution's ability to resist perturbations in which the amplitude is larger than the linear limit covered by a variational equation such as the one in Eq.~(\ref{Eq:variational}). This ability is generally captured by the size of the solution's basins of attraction. For the synchronized solution, we apply such perturbations at the initial instant of time by starting one network unit outside the synchronization manifold ${\bf s}$. Specifically, the oscillators on the synchronization manifold are initialized at the coordinates $(x_{S}^{t=0},y_{S}^{t=0})$, while the perturbed unit starts at $(x_{p}^{t=0},y_{p}^{t=0})$. To assess the effects of such perturbation on the spatiotemporal dynamics of our network, we define an order parameter as:
\begin{equation}
\pazocal{Z} = \frac{1}{N} \sum\limits_{i=1}^N z_i,\qquad
z_i = \left\{
\begin{array}{lcl}
0, & \mbox{ for } & D_i \ge \delta,\\[2mm]
1, & \mbox{ for } & D_i < \delta,
\end{array}
\right.
\label{Eq:order_parameter}
\end{equation}
where $D_i$ is the measure of local order around the $i$th oscillator, which is given by
\begin{equation}
D_i = \frac{1}{2 M + 1} \sum\limits_{k=-M}^M \| \mathbf{r}_{i+k} - \overline{\mathbf{r}}_i \|,
\label{Eq:local_order}
\end{equation}
with:
\begin{equation}
\overline{\mathbf{r}}_i = \frac{1}{2 M + 1} \sum\limits_{k=-M}^M \mathbf{r}_{i+k}.
\label{Eq:average}
\end{equation}
The parameters $M$ and $\delta$ are the number of neighbors and the tolerance considered in estimating the local synchrony around the $i$th oscillator. These values are fixed throughout the study at $M=2$ and $\delta=0.1$. With this, after the time evolution of the perturbed network for $t_{end}=1200T$, we evaluate the order parameter $\pazocal{Z}$. For $\pazocal{Z}=1$, we conclude that the completely synchronized state has been restored; otherwise, for $\pazocal{Z}<1$, the network is assumed to be in an alternative state. We distinguish two possible alternative states, namely: the completely desynchronized state for $\pazocal{Z}=0$ and the chimera state for $0<\pazocal{Z}<1$.

To estimate the global stability of the synchronized state $s$, we invoke the concept of single-node basin stability \cite{Mitra2017}. More specifically, we perform a number $I$ of different realizations of the system in Eq. (\ref{Eq:x_explicit}), each with a different initial condition $(x_{p}^{t=0},y_{p}^{t=0})$ attributed to the perturbed unit. These initial conditions are randomly taken in the interval $-8.0<x_{p}^{t=0}<8.0$ and $-20.0<y_{p}^{t=0}<40.0$. The initial conditions $(x_{S}^{t=0},y_{S}^{t=0})$ attributed to synchronization manifold are fixed at $(-5.0,5.0)$. Subsequently, for each realization, we evaluate the order parameter $\pazocal{Z}$, counting the number of realizations $I_S$ in which the network completely synchronizes ($\pazocal{Z}=1$). Therefore, the probability of synchronization $P_S=I_S/I$ yields an approximation of the relative size of the synchronization basin. Similarly to linear stability, the global stability of the synchronization manifold shall also depend on the network parameters. Hence, for $I=100$ realizations, in Fig.~\ref{Fig:figure_2}(b), we show the probability of synchronization $P_S$ as a function of the coupling intensity $\sigma$ for three different values of the coupling range $R$. We observe the existence of an interval of $\sigma$ in which the probability of synchronization decreases depending on the considered values of $R$. This observation indicates the existence of alternative states competing with the completely synchronized one for the network state space.

\begin{figure}[!htp]
\centering
\includegraphics[width=8.5cm,height=3.23cm]{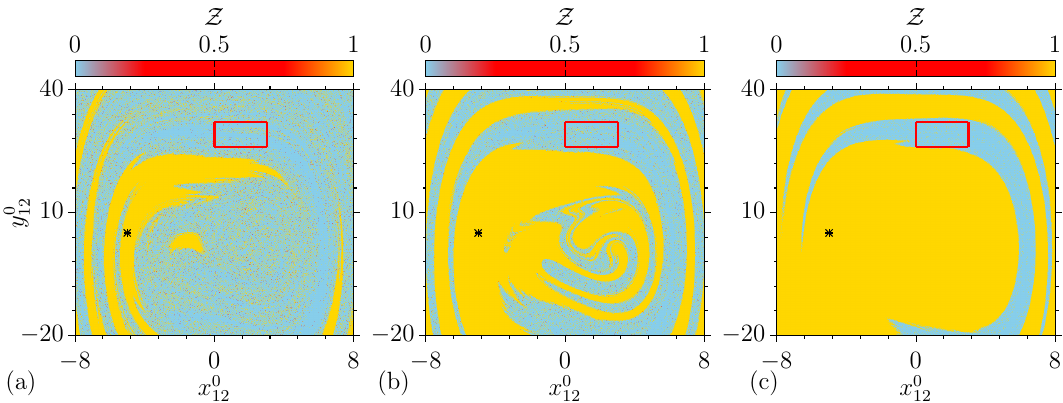}
\caption{Two-dimensional grid of initial conditions [($(x_{12}^{t=0},y_{12}^{t=0})$)] attributed to the perturbed unit $i=12$. Yellow corresponds to initial conditions leading to the completely synchronized state (synchronization basin), while light blue corresponds to initial conditions leading to the completely desynchronized state. The red dots correspond to transient chimera states. The coupling intensity is fixed at $\sigma=0.004$, and the network size is $N=25$. The coupling range is (a) $R=3$, (b) $R=5$, and (c) $R=7$. The black star marks the IC $(-5.0,5.0)$ attributed to the synchronized manifold, while the red square marks the interval $[0.0,3.0] \times [26.0,32.0]$ of ICs attributed to the perturbed unit.}
\label{Fig:figure_3}
\end{figure}

In order to better visualize such coexistences in the network state space, we obtain the order parameter $\pazocal{Z}$ for a grid of the initial conditions (ICs) ($(x_{p}^{t=0},y_{p}^{t=0})$) attributed to the perturbed oscillator. This procedure regards a two-dimensional cross-section of the network state space at the initial instant. Hence, for a network size fixed at $N=25$ and a coupling intensity set at $\sigma=0.004$ [vertical red line in Fig.~\ref{Fig:figure_2}(b)], we obtain this two-dimensional cross-section for the coupling range fixed at $R=3$, $R=5$, and $R=7$ in Figs.~\ref{Fig:figure_3}(a)-\ref{Fig:figure_3}(c), respectively. In these figures, yellow corresponds to the cross-section of the synchronization basin, i.e., the ICs leading to the completely synchronized state ($\pazocal{Z}=1$). The light blue color corresponds to ICs leading the network to complete desynchronization ($\pazocal{Z}=1$). The red-colored ICs correspond to transient network trajectories with $0<\pazocal{Z}<1$, therefore, possessing a coexistence of incoherent and coherent spatial domains. We regard these transitory spatiotemporal patterns as chimera states of our networks. As visible in Figs.~\ref{Fig:figure_3}(a)-Fig.~\ref{Fig:figure_3}(c), the area corresponding to the synchronization basin increases for increasing coupling range. This observation indicates that the completely desynchronized or chimera states depend strongly on the coupling range.

\section{Chimera states as consequence of trapping in nonattracting chaotic sets}
\label{Sec:Chimera_states}

\begin{figure}[!htp]
\centering
\includegraphics[width=8.5cm,height=6cm]{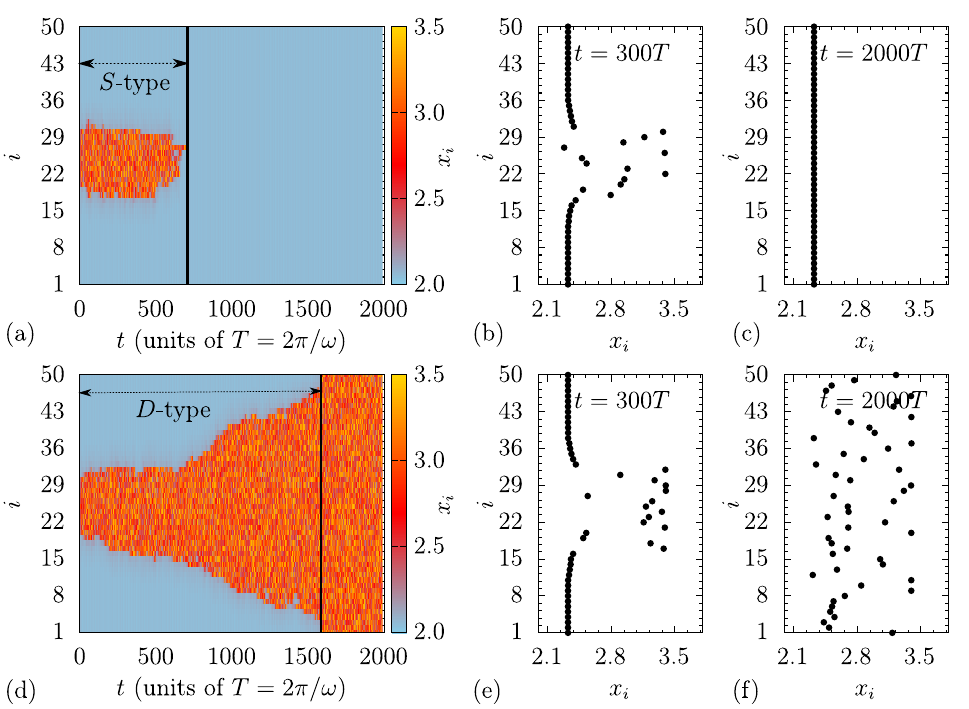}
\caption{(a)-(c) For the perturbed unit starting at $(-1.705,25.636)$: Spatiotemporal evolution of the network of Duffing oscillators, $S$-type chimera state, and the completely synchronized state, respectively. (d)-(f) For the perturbed unit starting at $(-1.656,25.557)$: Spatiotemporal evolution of the network of Duffing oscillators, $D$-type chimera state, and the completely desynchronized state, respectively. The space-time plots in (a) and (d) show only the time instants that are multiples of $3T$ where $T$ is the forcing period. The network parameters are $\sigma=0.004$, $N=50$, and $R=5$.}
\label{Fig:figure_4}
\end{figure}

We now inspect the dynamical features of the coexisting network solutions (shown in Fig.~\ref{Fig:figure_3}) and occurring for coupling intensities in the interval with a low probability of synchronization shown in Fig.~\ref{Fig:figure_2}(b). More specifically, we fix the coupling intensity at $\sigma=0.004$ [vertical red line in Fig.~\ref{Fig:figure_2}(b)] and study spatiotemporal evolution of the network for two realizations with different values of initial conditions attributed to the perturbed unit. Hence, for a network with $N=50$ oscillators, the oscillators in the synchronized manifold are initialized at $(-5,5)$ for both realizations. First, in Figs. \ref{Fig:figure_4}(a)-\ref{Fig:figure_4}(c), we initialize the perturbed unit $i=25$ at $(-1.705,25.636)$. The corresponding spatiotemporal evolution shows an initial phase of motion in which a coherent and an incoherent spatial domain coexist, forming a chimera state [Fig. \ref{Fig:figure_4}(a)]. A snapshot of the spatial configuration at $t=300T$ illustrates this chimera state [Fig. \ref{Fig:figure_4}(b)]. The incoherent spatial domain of this chimera state terminates abruptly at $t \approx 700T$ giving place to a completely synchronized state visible in the snapshot at $t=2000T$ [Fig. \ref{Fig:figure_4}(c)]. We denote the chimera states with this termination mechanism as $S$-type. Next, in Figs. \ref{Fig:figure_4}(d)-\ref{Fig:figure_4}(f), we initialize the perturbed unit $i=25$ at a different initial condition, specifically $(-1.656,25.557)$. The corresponding spatiotemporal evolution in Fig. \ref{Fig:figure_4}(d) also demonstrates the existence of a chimera state which is confirmed by the snapshot at $t=300T$ shown in Fig. \ref{Fig:figure_4}(e). However, in contrast to the previous case, in Fig. \ref{Fig:figure_4}(a), the incoherent spatial domain of this chimera state grows in size, eventually comprising the whole network at $t \approx 1580T$. The resulting completely desynchronized state is visible in the snapshot at $t=2000T$ [Fig. \ref{Fig:figure_4}(f)]. We denote the chimera states terminating this way $D$-type.

\begin{figure}[!htp]
\centering
\includegraphics[width=4.5cm,height=4.cm]{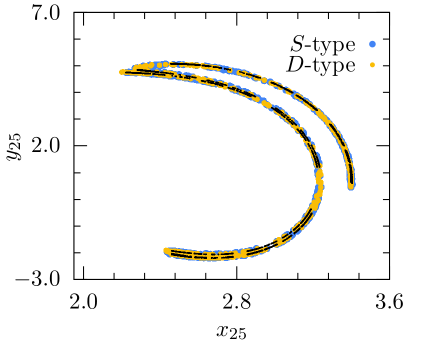}
\caption{The black dots approximate the chaotic saddle $\Lambda$ occurring in individual Duffing oscillators. The circles represent the trajectory of the network oscillator $i=25$ in the time interval $20T<t<600T$ occupying: The $S$-type chimera state (blue circles). The $D$-type chimera state (yellow circles). The network size and coupling intensity are $N=50$ and $\sigma=0.004$, respectively.}
\label{Fig:figure_5}
\end{figure}

Interestingly, the incoherent domain of both chimera states is shown in Figs. \ref{Fig:figure_4}(a) and \ref{Fig:figure_4}(d) correspond to oscillator trajectories trapped in the vicinity of the chaotic saddle $\Lambda'$ in the high-dimensional state space which -- projected to the $(x_{25},y_{25})$ plane -- looks similar to the chaotic saddle $\Lambda$ occurring for the isolated oscillators. This behavior can be visualized by inspecting the state space of specific oscillators in the incoherent spatial domain and comparing it with the chaotic saddle $\Lambda$ of isolated oscillators. For example, since the perturbed oscillator $i=25$ is at the center of the incoherent spatial domain of the network, we compare its trajectory $(x_{25},y_{25})$ in the time interval $20T<t<600T$ with the coordinates of the chaotic saddle $\Lambda$ [black dots in Fig.~\ref{Fig:figure_1}(a)]. Hence, in Fig.~\ref{Fig:figure_5}, we observe that the trajectory of the oscillator $i=25$ occupying the $S$-type (blue circles) and $D$-type (yellow circles) chimera states overlap with the chaotic saddle $\Lambda$ (black dots). These observations confirm that both chimera states observed in Figs. \ref{Fig:figure_4}(a) and \ref{Fig:figure_4}(d) are the result of the trapping of network oscillators in the chaotic saddle. The mechanism giving rise to this phenomenon has been explained in Ref. \cite{Medeiros2018} to justify the existence of the completely desynchronized state.

\begin{figure}[!htp]
\centering
\includegraphics[width=8.5cm,height=2.8cm]{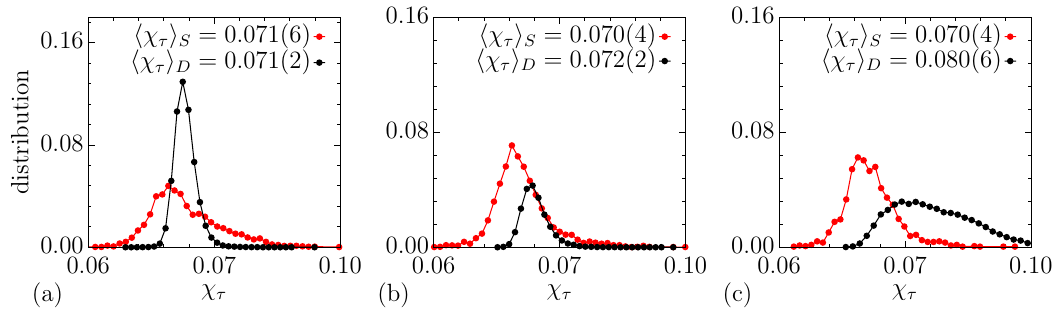}
\caption{Probability distribution of the largest Lyapunov exponent $\chi_{\tau}$ estimated during the chimera's lifetime $\tau$ for coupling ranges: (a) $R=3$. (b) $R=5$. (c) $R=7$. Red dots indicate $\chi_{\tau}$ corresponding to $S$-type chimeras, while black dots represent $D$-type chimeras. The network size and coupling intensity are $N=25$ and $\sigma=0.004$, respectively.}
\label{Fig:figure_6}
\end{figure}

In order to further illustrate the characteristics of transient dynamics resulting from trajectories trapped in the high-dimensional nonattracting chaotic set $\Lambda'$, we estimate the largest Lyapunov exponent (LLE) \cite{Wolf1985} of network trajectories during the chimera's lifetime. To accomplish this task, we consider ensembles of the ICs ($(x_{p}^{t=0},y_{p}^{t=0})$) attributed to the perturbed oscillator randomly taken in the intervals $x_{p}^{t=0} \in [0.0,3.0]$ and $y_{p}^{t=0} \in [26.0,32.0]$ [see the red square in Fig.~\ref{Fig:figure_3}]. In addition, the oscillators in the synchronized manifold are initialized at $(-5,5)$. For each trajectory, we estimate the LLE $\chi_{\tau}$ for the duration of the chimera's lifetime, denoted by $\tau$. Hence, in Figs.~\ref{Fig:figure_6}(a)-(c), we depict the probability distribution of $\chi_{\tau}$ for $S$-type (red dots) and $D$-type (black dots) chimeras for the previously considered values of the coupling range, namely, $R=3$, $R=5$, and $R=7$, respectively. We observe positive mean values of $\chi_{\tau}$ for $S$-type ($\langle \chi_{\tau} \rangle_S$) and $D$-type ($\langle \chi_{\tau} \rangle_D$) chimera types for the three values of the coupling range, confirming the chaoticity of the transient dynamics. In addition, for the three values of $R$ shown in Fig.~\ref{Fig:figure_6}, the mean values of $\chi_{\tau}$ for $S$- and $D$-type fall within the same confidence interval. This feature indicates the underlying role of the chaotic saddle $\Lambda'$ to the dynamics of the chimeras patterns observed here.

\begin{figure}[!htp]
\centering
\includegraphics[width=8.5cm,height=3.7cm]{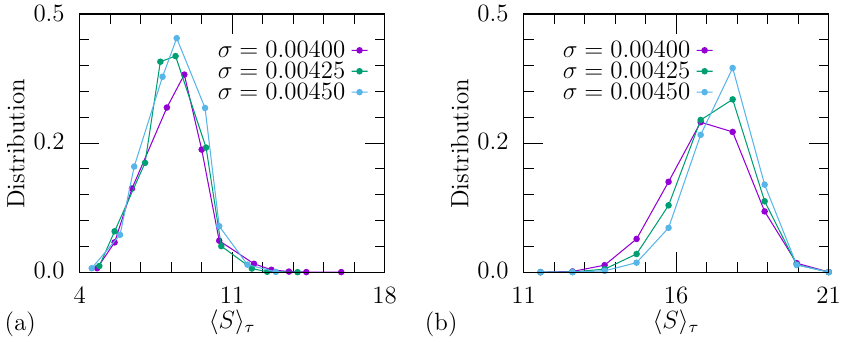}
\caption{Probability distribution of the temporal average of the size of the chimera's incoherent domain, $\langle S \rangle_{\tau}$ for three values of the coupling intensity $\sigma$. (a) For $S$-type and (b) for $D$-type chimeras. The network size and coupling range are $N=25$ and $R=3$, respectively.}
\label{Fig:figure_7}
\end{figure}

Another interesting aspect of the chimera states reported here is the movement of the interface between the coherent and incoherent domains. As visible in the spatiotemporal diagrams of Fig.~\ref{Fig:figure_4}, the size of the incoherent domain oscillates irregularly for both chimera types. To analyze this feature, we first consider the size of the incoherent domain to be a function of time, $S(t)$. Next, we take the temporal average of $S(t)$ over the chimera lifetime, $\tau$, to obtain the average chimera size $\langle S \rangle_{\tau}$. This procedure is repeated for ensembles of trajectories selected as in Fig.~\ref{Fig:figure_6}. Hence, in Figs.~\ref{Fig:figure_7}(a)-(b), we show $\langle S \rangle_{\tau}$ for slightly different values of the coupling intensity $\sigma$ for $S$- and $D$-type chimera states, respectively. The coupling range is kept fixed at $R=3$ for this analysis. For the $S$-type chimeras, using the protocol defined in Eq.~(\ref{Eq:local_order}) to estimate $S(t)$, we observe the mean values of the size of the incoherent domain around $\langle S \rangle_{\tau} \approx 8.0$ for all considered values of $\sigma$ in Fig.~\ref{Fig:figure_7}. On the other hand, for the $D$-type chimeras, the mean values of the incoherent domain are around $\langle S \rangle_{\tau} \approx 17.0$. This difference is attributed to the growing character of the incoherent domain of the $D$-type chimeras, yielding the larger mean value. A deeper investigation of the interaction between the coherent and incoherent domains is provided in the next section.

\section{Interaction between chimera's coherent and incoherent domains}

\begin{figure}[!htp]
\centering
\includegraphics[width=8.5cm,height=4.5cm]{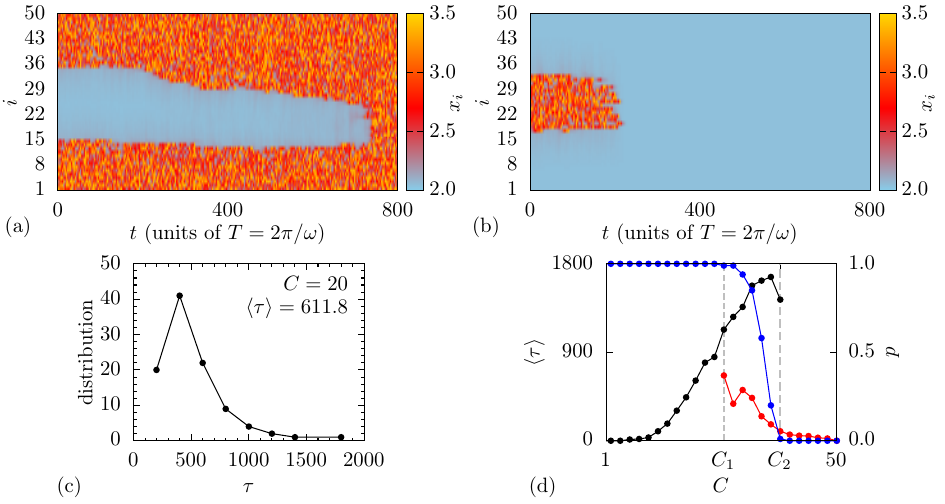}
\caption{(a)-(b) Spatiotemporal evolutions of $D$- and $S$-type chimera states obtained for ICs with the synchronized domain of size $C = 20$ and $C = 34$, respectively. (c) Distribution of $D$-type chimeras lifetimes for $100$ realizations with the synchronized domain of size $C = 20$. (d) Mean lifetimes of $D$-type chimeras (black) and $S$-type chimeras (red) as a function of the synchronized domain size $C$. The blue curve shows the fraction $p$ of ICs converging to the completely desynchronized state. The network parameters are $\sigma=0.004$, $N = 50$, and $R = 10$.}
\label{Fig:figure_8}
\end{figure}

For the system parameters used in the previous section, we observe that the completely synchronized state stably coexists with the completely desynchronized one. Such coexistence raises the question of whether interfaces can be formed connecting these two homogeneous states and what their properties are. In order to consider this issue, we perform a series of numerical simulations with specially designed ICs. To obtain such ICs, we first take a snapshot of the completely desynchronized state. Subsequently, we reset the values of the oscillators with indices $1$ to $C$ to the values of the completely synchronized state. This way, we obtain ICs consisting of coherent and incoherent domains of sizes $C$ and $N - C$, respectively. We emphasize that the ICs of the synchronized state lie on the period-3 attractor, $\mathbf{A}$, occurring in the state space of each oscillator. This protocol contrasts with the synchronized manifold receiving $(-5,5)$ as IC in the previous figures, yielding different sensitivity to perturbing a single unit. Next, starting numerical simulations from these ICs, we observe the following dynamical behaviors, see Figure~\ref{Fig:figure_8}.
\smallskip

(i) For $C \le R$, the network trajectories quickly collapse to the desynchronized state.
\smallskip

(ii) For $R < C < C_1$, where $C_1$ is a critical value, only the formation of $D$-type chimera states are observed, Fig.~\ref{Fig:figure_8}(a). This chimera state looks like a coherent region bounded by two non-monotonically moving fronts, which exhibit a long-term tendency to invade the coherent region from the incoherent one. Due to a certain degree of chaoticity in the movement of the fronts, the size of the coherent region could increase or decrease at different time intervals. Still, sooner or later, it decreases to the value of the network coupling range $R$, and then the collapse of the chimera state into the desynchronized state occurs. The time from the start of the simulation to collapse corresponds to the lifetime $\tau$ of the chimera, and it takes different values for different realizations of the completely desynchronized state. In particular, using $100$ independent realizations of the desynchronized state and generating corresponding ICs for each $C$, we calculate the distribution of lifetimes from which we obtain the mean lifetime $\langle\tau\rangle$ (see the histogram in Fig.~\ref{Fig:figure_8}(c) for $C=20$). The dependence of the mean lifetime $\langle\tau\rangle$ of the $D$-type chimeras on the size of the coherent section $C$ is shown by the black curve in Fig.~\ref{Fig:figure_8}(d).
\smallskip

(iii) For $C_1 \le C \le C_2$, where $C_2$ is another critical value, bistability occurs in the system. Some ICs continue converging to the $D$-type chimera states, while others converge to the $S$-type chimeras, Fig.~\ref{Fig:figure_8}(b). The probability of getting a $D$-type chimera state is a decreasing function of $C$ tending to $0$ for $C\to C_2$ (see blue curve in Fig.~\ref{Fig:figure_8}(d)).
\smallskip

(iv) For $C > C_2$, all ICs lead to $S$-type chimeras, while $D$-type chimeras are no longer observed. This observation contrasts the results presented in Section~\ref{Sec:Chimera_states}, where bistability occurs by perturbing a single unit. Such difference is due to the $C$ oscillators composing the coherent domain of the chimera being initialized very close to the period-$3$ attractor in this protocol, reducing the sensitivity of synchronized states to perturbations of a single unit. Moreover, the mean lifetime of $S$-type chimeras decreases in the interval $C \ge C_1$ (see the red curve in Fig.~\ref{Fig:figure_8}(d)).
\smallskip

Most of the above features are found in other simulations with different values of $N$, $R$, and $\sigma$ for which $D$-type and $S$-type chimeras can occur. Importantly, the critical values $C_1$ and $C_2$ show a nonlinear dependence on the system parameters. We leave its analysis for future investigations.

%%it is counterintuitive that you increase C and get larger mean lifetimes of D-type chimeras.
%%in item iii) there was a statement about S-type chimeras where it should be D-type.
\section{Lifetime of the chimera states}

\begin{figure}[!htp]
\centering
\includegraphics[width=8.5cm,height=5.5cm]{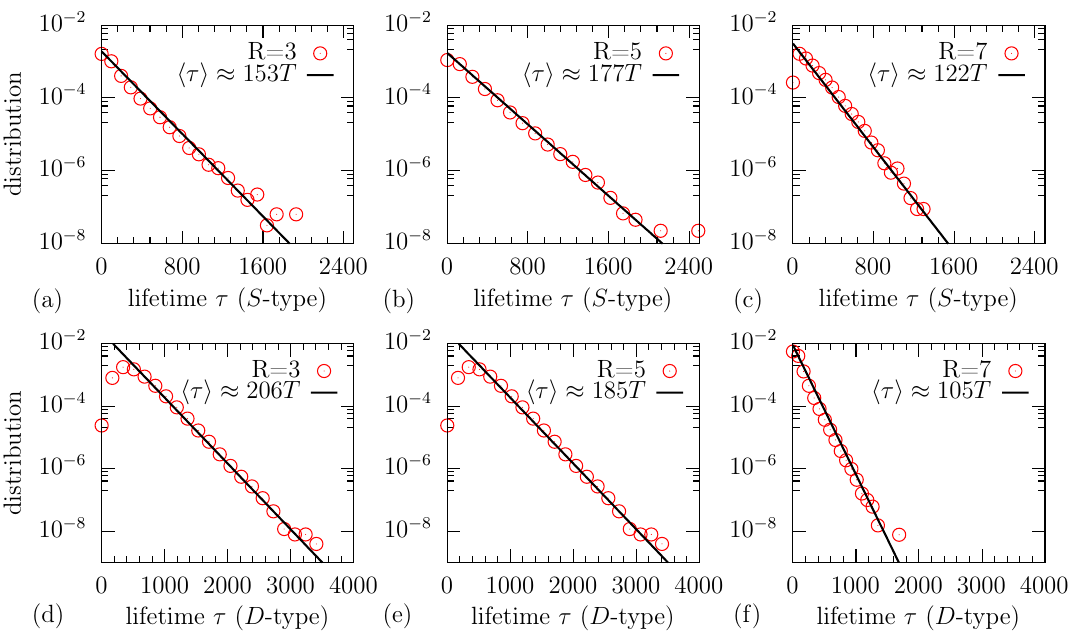}
\caption{(a)-(c) Probability distribution of lifetimes $\tau$ of $S$-type chimera states for the coupling range $R=3$, $R=5$, and $R=7$, respectively. (d)-(f) Similarly, the distribution of $\tau$ of $D$-type chimera states for the three values of coupling range. The mean lifetime $\langle \tau \rangle$ is obtained from an exponential regression (black curve) of the observed lifetimes (red circles). The network size and coupling intensity are $N=25$ and $\sigma=0.004$, respectively.}
\label{Fig:figure_9}
\end{figure}

In the literature, the chimera states have often been regarded as chaotic transients for networks of phase oscillators \cite{Wolfrum2011}. In such context, the chimera's lifetime has been shown to increase exponentially for increasing system size \cite{Wolfrum2011,Lilienkamp2020}. We expect that an increasing system size in Eq.~(\ref{Eq:x_explicit}) would lead to large mean escape times of the high-dimensional chaotic saddle $\Lambda'$ \cite{Crutchfield1988, Lai1995,Lai2011}. To verify if this is also the case for networks of nonlinear oscillators possessing nonattracting chaotic sets in their state space, we now investigate the lifetime of the chimera states observed here for different coupling ranges and sizes. Hence, for the network of Duffing oscillators, as described in Section~\ref{Sec:Chimera_states}, we perform a statistical analysis over ensembles of trajectories starting with different values of ICs ($(x_{p}^{t=0},y_{p}^{t=0})$) attributed to the perturbed oscillator. These ICs are randomly taken in the intervals $x_{p}^{t=0} \in [0.0,3.0]$ and $y_{p}^{t=0} \in [26.0,32.0]$ [see the red square in Fig.~\ref{Fig:figure_3}]. The oscillators in the synchronized manifold are initialized at $(-5,5)$ [see the black star in Fig.~\ref{Fig:figure_3}]. We investigate the lifetime $\tau$ of the chimera states for different values of the coupling range $R$, keeping the network size and coupling intensity fixed at $N=25$ and $\sigma=0.004$, respectively. In Figs.~\ref{Fig:figure_9}(a)-\ref{Fig:figure_9}(c), we show the distribution of lifetimes $\tau$ of $S$-type chimera states for $R=3$, $R=5$, and $R=7$, respectively. Similarly, in Figs.~\ref{Fig:figure_9}(d)-\ref{Fig:figure_9}(f), we show the distribution of lifetimes of $D$-type chimera states for the three different values of the coupling range $R$. In these figures, we first verify that the lifetime of both types of chimera states is exponentially distributed --- evidence for a high-dimensional nonattracting chaotic set mediating the lifetime of the chimera states. The mean lifetime $\langle \tau \rangle$ is obtained from an exponential regression $\rho(\tau) =\rho_0 e^{-\tau/\langle \tau \rangle}$ of the observed data. The values of $\langle \tau \rangle$ are shown individually in Figs.~\ref{Fig:figure_9}(a)-\ref{Fig:figure_9}(f) for each value of $R$. Interestingly, for the $D$-type chimera states, we observe a monotonic decrease of their mean lifetime $\langle \tau \rangle$ as the coupling range increases. However, for the $S$-type chimera states, the mean lifetime is larger for the intermediate coupling range values $R=5$. This observation raises the question about the existence of an optimal coupling range for longer lifetimes of the chimera states studied here.

\begin{figure}[!htp]
\centering
\includegraphics[width=8.5cm,height=4cm]{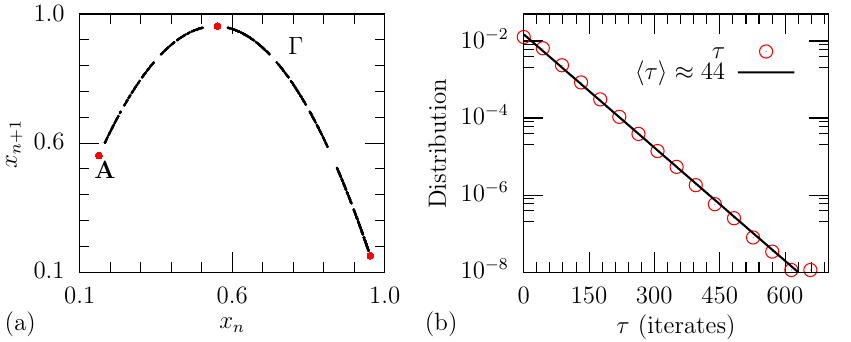}
\caption{(a) State space of the logistic map ($\sigma=0$). The red circles represent the period-$3$ orbit, while the black dots display an approximation of the chaotic repeller. (b) The red circles indicate the distribution of escape times of the chaotic repeller. The black curve is an exponential fit providing the mean escape time $\langle \tau \rangle \approx 44$ map iterates.}
\label{Fig:figure_10}
\end{figure}

To better address this question, we recall that the essential ingredients for the onset of our chimera states are very common, i.e., a nonattracting chaotic set and a stable periodic orbit. Therefore, these chimera states can appear in networks of different systems. We take advantage of this fact to utilize systems that are easier to simulate, facilitating the study of more extensive networks and longer evolution times. Therefore, we now consider networks of discrete-time units possessing the two ingredients of interest. The high-dimensional equation describing such a system is given by:
\begin{equation}
    x_{i}^{t+1}  = f(x_i^t) + \frac{\sigma}{2R}\sum_{j=i-R}^{j=i+R}[f(x_{j}^{t})-f(x_{i}^{t})],
    \label{logistic}
\end{equation}
where the function $f(x)$ is defined to be the logistic map, i.e., $f(x)=ax(1-x)$. For the parameter $a=3.8205$, the logistic map oscillates in a stable period-$3$ attractor, which coexists with a chaotic repeller. For $\sigma=0$, this coexistence is shown in Fig.~\ref{Fig:figure_10}(a) where the period-$3$ attractor $\mathbf{A}$ is indicated by red circles and an approximation of the chaotic repeller $\Gamma$ is shown in black. In Fig.~\ref{Fig:figure_10}(b), we show the chaotic repeller's respective distribution of escape times.

\begin{figure}[!htp]
\centering
\includegraphics[width=8.5cm,height=6.5cm]{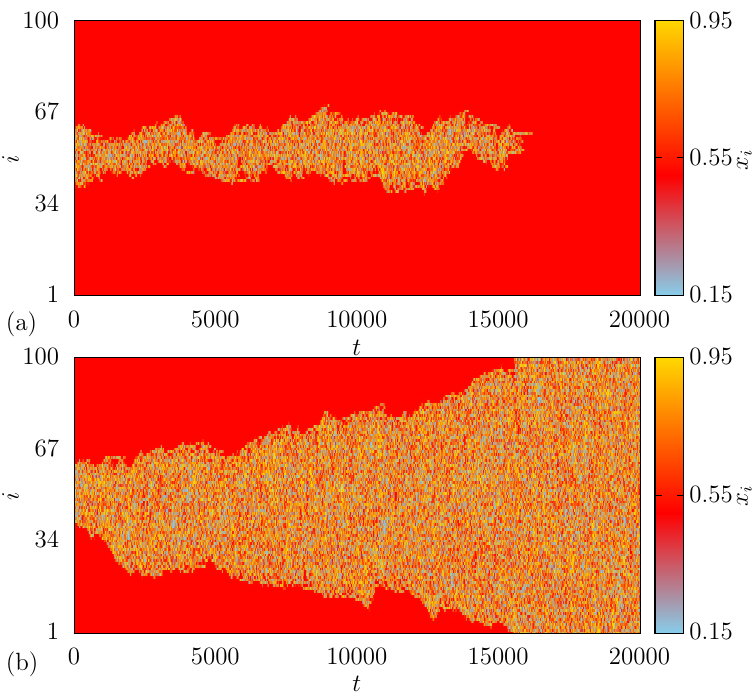}
\caption{Spatiotemporal diagram of the network of logistic maps as units (Eq.~(\ref{logistic})). For the perturbed unit starting at: a) $x_{p}^{t=0}=0.26$ leading the network to a $S$-type chimera state. b) $x_{p}^{t=0}=0.90$ leading the network to a $D$-type chimera state. The network size and coupling radius are $N=100$ and $R=5$, respectively. These space-time plots show $x_i$ only at every three map iterates $t$. The coupling intensity is $\sigma=0.001$.}
\label{Fig:figure_11}
\end{figure}

We now consider the coupled system [$\sigma \neq 0$ in Eq.~(\ref{logistic})] to illustrate the occurrence of $S$-type and $D$-type chimera states. We first emphasize that the network coupling introduces a stable direction to the chaotic repellers in the one-dimensional state space of the logistic map, transforming the chaotic repellers $\Gamma$ of individual maps into a high-dimensional chaotic saddle $\Gamma'$ of the network. Next, we perturb the synchronized state at $t=0$ by initializing the maps in the synchronization manifold at the initial condition $x_{S}^{t=0}=0.3$ and one unit (perturbed unit) at a different initial condition represented by $x_{p}^{t=0}$. For $N=100$ and $R=5$, in Fig.~\ref{Fig:figure_11}, we show the time evolution for two different values of ICs attributed to the perturbed unit. For $x_{p}^{t=0}=0.26$, we observe the occurrence of both types of chimera states, $S$-type [Fig.~\ref{Fig:figure_11}(a)] and $D$-type [Fig.~\ref{Fig:figure_11}(b)] for $x_{p}^{t=0}=0.90$. Similarly to the network of Duffing oscillators, these chimera states are also a consequence of trapping trajectories in the chaotic saddle $\Gamma'$ occurring in the high-dimensional state space of the network.

\begin{figure}[!htp]
\centering
\includegraphics[width=8cm,height=8cm]{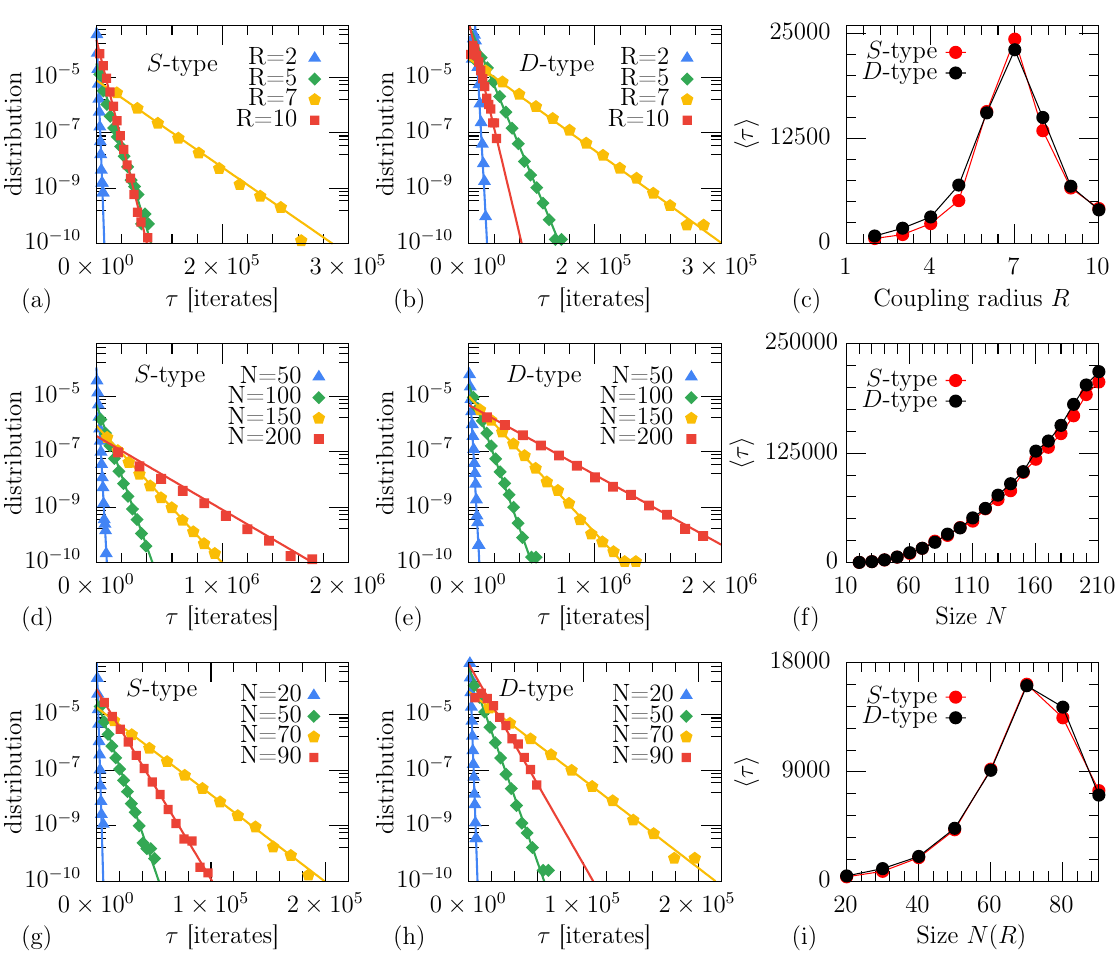}
\caption{For the network of logistic maps in Eq.~(\ref{logistic}) with coupling intensity fixed at $\sigma=0.001$. (a)-(b) Respective normalized probability distribution of lifetimes $\tau$ of $S$-type and $D$-type chimera states for four values of the coupling range $R$ (colors). (c) Mean chimera lifetime $\langle \tau \rangle$ as a function of $R$. The network size is fixed at $N=80$. (d)-(e) Respective normalized probability distribution of both chimera types for different values of the network size $N$ (colors). (f) Mean lifetime $\langle \tau \rangle$ as a function of $N$. The coupling range is fixed at $R=7$. For $R=0.1\times N$ in (g)-(h) are the respective normalized probability distribution of lifetimes for $S$-type and $D$-type chimera states for different $N$. (i) Mean lifetime $\langle \tau \rangle$ as function of $N(R)$.}
\label{Fig:figure_12}
\end{figure}

Following the verification of the occurrence of both chimera states in the network of logistic maps, we now investigate the dependence of their lifetime on the coupling range $R$ and size $N$. For this purpose, we consider ensembles of trajectories starting with different values of ICs $(x_{p}^{t=0}$ randomly taken in the interval $x_{p}^{t=0} \in [0.1,0.85]$. The coupling intensity is fixed at $\sigma=0.001$ throughout the entire analysis of this network. Hence, we first examine the normalized distribution of lifetimes of $S$-type chimera states [Fig.~\ref{Fig:figure_12}(a)] and $D$-type chimera states [in Fig.~\ref{Fig:figure_12}(b)] for different network coupling ranges $R$ indicated by the different colors in these figures. The network size is held constant at $N=80$. Similarly to the network of Duffing oscillators, we observe an exponential probability distribution of the chimeras lifetime $\tau$. We obtain the mean lifetimes $\langle \tau \rangle$ via an exponential data regression. As a result,  in Fig.~\ref{Fig:figure_12}(c), we show the corresponding $\langle \tau \rangle$ as a function of the coupling range $R$. As suspected from our results for the network of Duffing oscillators, we observe here the existence of an optimal coupling range yielding the longest lifetimes for both chimera states. Subsequently, for the coupling range fixed at the optimal value $R=7$, we study the distribution of lifetimes of the chimera states for different network sizes indicated by the different colors in Fig.~\ref{Fig:figure_12}(d) and Fig.~\ref{Fig:figure_12}(e). The mean lifetime $\langle \tau \rangle$ as a function of $N$ is shown in Fig.~\ref{Fig:figure_12}(f). At this configuration, we verify that the lifetimes increase monotonically within the considered interval of $N$. This observation agrees with findings in which the mean escape time from the chaotic saddle increases with increasing system size \cite{Crutchfield1988, Lai1995}. Next, we consider that the previous works \cite{Wolfrum2011,Lilienkamp2020} investigating the lifetime of chimera states consider a linear dependence of the coupling range $R$ on the network size $N$ as $R=r \times N$. The parameter $r$ is kept constant, maintaining proportionality between $R$ and $N$ for increasing values of $N$. In this context, the lifetime of chimera states and chaotic transients are generally known to increase fast with the system size \cite{Wolfrum2011,Lilienkamp2020}. We now test this possibility in our chimera states by fixing $r$ at $0.1$ and varying $N$. In Figs.~\ref{Fig:figure_12}(g) and \ref{Fig:figure_12}(h), we show the normalized probability distribution of lifetimes for $S$-type and $D$-type chimera states, respectively. Similarly to the previous scenarios, the lifetimes are exponentially distributed. Following an exponential regression of the data ($\rho(\tau) =\rho_0 e^{-\tau/\langle \tau \rangle}$), we obtain the mean lifetimes $\langle \tau \rangle$ for an interval of $N$. We observe that $\langle \tau \rangle$ grows to a network size $N=70$ corresponding to the optimal coupling range $R=7$.

\section{Conclusions}

In summary, we report a new chimera state formation mechanism based on trapping network trajectories in nonattracting chaotic sets present in the state space of each network unit. More specifically, in this study, we consider two different nonlinear systems as the units of our networks: Duffing oscillators in which the nonattracting chaotic set is a chaotic saddle and logistic maps with a chaotic repeller. When each of those systems is taken as the dynamics of a unit in the network, we obtain, in both cases, a chaotic saddle in the high-dimensional state space, which is the main cause for the formation of the chimera states reported here. Moreover, the nonattracting chaotic sets of both classes of units coexist with stable periodic orbits. In this scenario, we find that a subset of the network units is trapped near the nonattracting chaotic set forming the incoherent spatial domain of chimera states. On the other hand, the complement subset of units is in the stable periodic orbit creating the coherent spatial domain of the chimera state.

In addition, we find that the termination of the chimera states in these networks occurs in two different ways:

(i) The incoherent domain of the chimera state abruptly terminates, and the network completely synchronizes. We denote the chimera states with this termination scheme $S$-type.
The linear stability of the completely synchronized state for all positive coupling intensities of the considered network of Duffing oscillators is demonstrated with the formalism of the master stability function.

(ii) The incoherent domain of the chimera states continuously grows, resulting in an entirely desynchronized network. We denote the chimera states with this termination scheme $D$-type.

For both types of chimera states, their lifetimes have an exponential distribution (but with slightly different parameters), typical of chaotic transients. By analyzing the dependence of the chimera states on the network parameters (coupling range and size), we find that the lifetimes of chimera states of $S$-type depend non-monotonically on the coupling range in a network of Duffing oscillators. We further investigate this observation in the network of logistic maps,
confirming the occurrence of an optimal coupling range for the lifetime of both chimera states. Concerning the dependence on the network size, we observe a monotonic growth of the chimera's lifetime as the network size increases for a fixed coupling range. However, considering the coupling range as a fraction of the network size, the longest lifetimes are obtained for an optimal network size, reflecting the dependence on the coupling range.

\begin{acknowledgments}
E.S.M and U.F. acknowledge the support by the Deutsche Forschungsgemeinschaft (DFG) via the project number 454054251 (FE 359/22-1). The simulations were performed at the HPC Cluster CARL, located at the University of Oldenburg (Germany) and funded by the DFG through its Major Research Instrumentation Program (INST 184/157-1 FUGG) and the Ministry of Science and Culture (MWK) of the Lower Saxony State, Germany. The work of O.E.O. was supported by the Deutsche Forschungsgemeinschaft under Grant No. OM 99/2-2.
\end{acknowledgments}

\section*{Data Availability Statement}

The data that support the findings of this study are available from the corresponding author upon reasonable request.

\nocite{*}
% \bibliography{chimera}
%

\end{document}